# Optimization of Transferable Interatomic Potentials for Glasses toward Experimental Properties

Ruoxia Chen[a], Kai Yang[b], Morten M. Smedskjaer[c], N. M. Anoop Krishnan[d,e], Jaime Marian[f], Fabian Rosner[g,]*

a. Department of Civil and Environmental Engineering, University of California, Los Angeles, CA, USA, ruoxia@g.ucla.edu
b. Department of Civil and Environmental Engineering, University of California, Los Angeles, CA, USA, tzyk4707@outlook.com
c. Department of Chemistry and Bioscience, Aalborg University, Aalborg, Denmark, mos@bio.aau.dk
d. Department of Civil Engineering, Indian Institute of Technology Delhi, New Delhi, India, krishnan@iitd.ac.in
e. Yardi School of Artificial Intelligence, Indian Institute of Technology Delhi, New Delhi, India, krishnan@iitd.ac.in
f. Department of Materials Science and Engineering, University of California, Los Angeles, CA, USA, jmarian@g.ucla.edu
g. Department of Civil and Environmental Engineering, University of California, Los Angeles, CA, USA, fabianrosner@g.ucla.edu

*Corresponding authors: Fabian Rosner: fabianrosner@g.ucla.edu

## Abstract

The accuracy of molecular simulations is fundamentally limited by the interatomic potentials that govern atomic interactions. Traditional potential development, which relies heavily on *ab initio* calculations, frequently struggles to reproduce the experimentally observed properties that govern real material behavior. To address this challenge, we present a machine learning-driven, active-learning optimization framework for optimizing classical interatomic potentials to reproduce experimental properties. Our method, here showcased on soda-lime borosilicate glasses, targets both global (density) and local (boron coordination) structural features across a wide range of compositions. By combining a surrogate model with iterative active learning, the framework efficiently explores a five-dimensional parameter space using only 400 molecular dynamics simulations over 17 iterations, making it highly data-efficient and eliminating the need for extensive simulation campaigns. Two transferable parameter sets are identified, each demonstrating good agreement with experimental measurements, including glass density, fraction of four-fold boron, and X-ray structure factor. The framework effectively captures and manages inherent trade-offs between structural objectives and compositional regimes, providing insights into the coordination behavior of boron in complex glass networks. The resulting classical force fields are generalizable and do not require reparameterization for individual compositions. Altogether, this work offers a scalable and experimentally grounded approach for developing transferable interatomic potentials suitable for a broad range of materials, including multi-component glass systems, and beyond.

## Highlights:

- Developed an active-learning framework to optimize interatomic potentials using experimental data, achieving generalization across seven soda-lime borosilicate glass compositions.
- Jointly optimized for density and boron coordination, yielding two transferable parameter sets that accurately capture density, $B^4$ fraction, and X-ray structure factor.
- Effectively balanced trade-offs between global and local structural targets across diverse compositions.
- Enable composition-independent simulations of complex multi-component glasses without reparameterization.

## 1. Introduction

The accuracy of atomistic simulations critically depends on the quality of the interatomic potential parameters[1]. These parameters define the underlying interactions that govern atomic behavior, and their accuracy directly impacts the predictive power of molecular dynamics (MD) simulations[2,3]. However, identifying suitable parameter values is often challenging, particularly for complex multi-component systems where multiple atomic species and bonding environments coexist[4,5]. In such systems, a single potential must capture a wide range of structural and thermodynamic properties, including coordination preferences, defect energetics, and mechanical response[6]. Recent studies based on MD simulations have highlighted that poor parameterization can lead not only to quantitative errors in properties like density or elastic moduli, but also to qualitatively incorrect predictions of atomic structure and dynamics[7,8]. Additionally, parametrization of potentials for glasses is particularly challenging, given the non-equilibrium nature of glasses and specifically their cooling rate dependence. As such, optimizing the potential to reproduce crystalline structures or families does not guarantee the meaningful reproduction of their glassy counterparts [9].

An example is borosilicate glasses, which are widely used for technologically critical applications, ranging from nuclear waste containment to chemical resistance and optical devices[10]. These versatile materials derive their unique properties from a complex atomic structure, where network formers such as boron and silicon interact with various modifier cations (e.g., $Na^+$, $Ca^{2+}$) in a disordered environment at different length scales[11]. One of the major challenges in simulating borosilicate glasses lies in developing accurate interatomic potentials that can capture this structural complexity across diverse compositions[12,13]. A central difficulty stems from the highly composition-dependent coordination behavior of boron atoms[14,15]. Experimental studies have shown that the ratio between threefold-coordinated ($B^3$) and fourfold-coordinated ($B^4$) boron varies nonlinearly with the relative amounts of network formers and modifiers, a phenomenon often referred to as the "boron anomaly"[11]. For instance, changes in the Si/B ratio or modifier oxide content can significantly shift the equilibrium between $B^3$ and $B^4$ environments, leading to substantial effects on macroscopic properties such as density, hardness, and glass transition temperature. Accurately capturing this behavior in simulations requires potential functions that are sensitive to subtle compositional changes in bonding and coordination[16,17].

In addition to compositional sensitivity, a second major challenge in modeling borosilicate glasses lies in the temperature and fictive temperature dependence of boron speciation[18]. Experimental glasses are typically cooled slowly, allowing the atomic structure to relax toward low-energy configurations[19,20]. In contrast, MD simulations rely on high cooling rates that are orders of magnitude faster ($10^9$–$10^{14}$ K/s), resulting in glasses that are effectively trapped at higher fictive temperatures. These rapid quench rates lead to more entropically dominated atomic configurations, particularly affecting the coordination environment of boron atoms. Because the fictive temperature governs the relative populations of $B^3$ and $B^4$, it serves as a key indicator of thermal history and strongly influences the final glass structure. Under fast cooling, the conversion of $BO_3$ units to $BO_4^-$ is often incomplete, causing MD-generated glasses to underpredict the experimentally observed $B^4$ fractions. Despite this limitation, MD simulations remain valuable because they capture essential structural trends and local environments observed in real glasses and allow systematic exploration of compositional and thermal effects at the atomic scale. Thus, the potential parameters should be optimized so that the simulated glass structures reproduce experimental properties, despite the differences in the thermal history[20].

*Ab initio* molecular dynamics (AIMD) and density functional theory (DFT)-based MD have been used to study borate and borosilicate glasses[21,22,23]. These first-principles approaches offer high accuracy by explicitly solving the electronic structure problem without relying on predefined empirical potentials. AIMD methods are particularly well-suited to capturing the interplay between ionic and covalent bonding, electronic polarization, and subtle local structural features such as bond angles and coordination environments[24]. For example, Tanaka et al. demonstrated that DFT-MD simulations could be used to train or directly benchmark classical force fields for complex multi-component glasses[25]. Similarly, Jabraoui et al. used Car–Parrinello MD to study interfacial chemistry in Ca–Na borosilicate systems, showing that AIMD can reproduce experimental trends such as $BO_4$ fractions more reliably than traditional rigid-ion models[26]. However, despite their accuracy, these methods face significant limitations. They are computationally intensive, typically restricted to small system sizes (hundreds of atoms) and short timescales (picoseconds), and often require unrealistically fast cooling rates that can produce structural artifacts such as overly broad bond angle distributions[2]. Additionally, limited configurational sampling and inherent approximations in exchange–correlation functionals may lead to discrepancies in medium-range order (e.g., ring structures) and vibrational properties when compared to experimental data.

To reduce the *computational* cost of *ab initio* simulations, many studies have focused on fitting classical interatomic potentials to *ab initio* reference data[27]. Common strategies include brute-force grid or random searches. However, the stochastic nature of random search may miss optimal parameter combinations, especially in large and complex search spaces. Grid search, while systematic, becomes prohibitively expensive in high-dimensional parameter spaces and may waste computational resources exploring unimportant dimensions[28]. To overcome these limitations, researchers have explored more efficient optimization strategies, including machine learning (ML)–based methods[29]. For example, Urata et al. developed a DeepMD potential for Li-borosilicate glass that accurately reproduced structural features such as four-fold coordinated boron and three-membered rings, outperforming traditional empirical potentials[30]. Despite these successes, such ML potentials are typically trained on data from a single composition and require extensive DFT datasets, limiting their transferability[31]. On the other end of the spectrum, pre-trained universal ML potentials aim for high transferability across diverse chemistries and compositions, but they often sacrifice accuracy and efficiency for any specific glass system compared to models trained on targeted datasets. Without composition-dependent terms, even advanced models often fail to capture key structural trends, such as the shift in the $BO_3/BO_4$ ratio across different chemistries and thermal histories. As a result, while these models may perform well on the compositions they are trained on, they typically generalize poorly to others, impeding their broader applicability and transferability. Moreover, classical force fields derived from ML or *ab initio* fitting remain computationally expensive to develop and are difficult to scale to the complexity of multi-component systems like borosilicate. Most ML-based approaches are trained on narrow compositional datasets, limiting their ability to represent the full structural diversity observed experimentally[32]. Although these methods offer a promising link between quantum-level accuracy and classical MD efficiency, challenges in data generation, transferability, and structural sensitivity continue to hinder their broader applicability in glass potential development.

Another strategy for improving generalizability is to develop empirical potentials with parameters that adapt to composition. For example, Kieu et al. and Inoue et al. introduced composition-dependent charges and B–O interaction parameters to accurately reproduce $BO_3/BO_4$ ratios in Na–B–Si glasses[16,33]. While these models effectively capture boron coordination trends by design, they are limited in transferability to systems with different modifiers or multi-component formulations, often requiring manual adjustment for each new glass family. Bertani et al. addressed this by proposing a "self-consistent" interatomic potential in which B–O force constants vary with two compositional ratios[34]. Their Bayesian-optimized model successfully reproduced densities, $BO_4$ fractions, coordination environments, and even NMR spectra across Na, Li, Ca, and Mg borosilicate series. However, these approaches are inherently complex, relying on composition-specific functional forms and numerous empirical constants. Moreover, empirical models depend strongly on the quality of reference data, which can vary due to differences in thermal history during sample preparation[35]. This variability can propagate into the fitted parameters, making it difficult to obtain a universally robust potential. To address these limitations, Wang et al. developed a transferable interatomic potential by fitting to experimental measurements of both density and $B^4$ fraction across multiple borosilicate compositions[36]. Although the model demonstrated promising accuracy, the use of an incorrect atomic mass for boron reduced its physical fidelity and limited its applicability [37]. Kai et al. employed the interatomic potential for borosilicate glasses derived from this active learning framework to span a wide compositional range. By targeting experimental densities and four-fold coordinated boron fractions with proper atomic mass, they demonstrate successful potential parameterization. Their detailed analysis provides further validation that our methodology can be effectively applied to MD potential development for reproducing experimental target properties[38]

In this paper, we propose a novel ML–driven optimization framework. To showcase and validate our approach, we calibrate the parameters of the Beest Kramer van Santen (BKS) potential which is a simple and widely used empirical model that effectively describes Si–O interactions in silicate systems but contains many composition-dependent parameters that require refinement. We use experimental data, specifically targeting density and average $B^4$ coordination across a wide range of soda-lime borosilicate glass compositions, to optimize these parameters. Compared to *ab initio* reference data, experimental measurements offer a more direct and accurate representation of real glass structure. To address the limitations of small datasets and reduce the reliance on computationally expensive simulations, our framework incorporates a data-efficient, active learning strategy that explicitly manages the trade-off between global (density) and local ($B^4$ fraction) structural properties. By iteratively refining an ML surrogate model trained on MD data and guiding the parameter search through active learning, our method efficiently explores the high-dimensional parameter space and identifies transferable parameter sets, that are transferable across a wide-range of compositions. This enables the resulting interatomic potential to achieve high accuracy across structurally diverse glass chemistries. Overall, our approach provides a robust and scalable pathway for developing classical potentials that are optimized to reproduce experimental data, facilitating realistic predictive simulations of complex multi-component glass systems and even other material systems.

## 2. Methods

**2.1 Potential parameter optimization framework**

To efficiently identify generalized parameters of the BKS potential suitable for a wide range of $Na_2O$–$CaO$–$B_2O_3$–$SiO_2$ glasses with varying compositions, we developed an active learning–driven framework that integrates MD simulations (see Section 2.2) with an ML surrogate model, that takes the potential parameters and composition as input, and predicts the $B^4$ fraction and density. Note that although only two experimentally measured properties are used in this case, the framework can, in principle, be extended to any number of experimental properties. Optimized parameter sets are obtained through a multi-objective optimization process, as illustrated in Figure 1. The key idea is to significantly reduce the number of expensive MD simulations by using a multi-layer perceptron (MLP) model to approximate simulation outputs and guide the parameter search.

We start by constructing an initial dataset from classical MD simulations using the BKS potential, which includes predicted density and $B^4$ fraction values for varying values of the BKS potential parameters (to be optimized), across multiple glass compositions. These data are then used to train an MLP model, that predicts the density and $B^4$ fraction values based on the potential parameters and the $B_2O_3$ molar fraction. The resulting composition-aware model can predict glass properties efficiently in terms of computational cost across seven different $B_2O_3$ and $SiO_2$ contents with constant $Na_2O$ and $CaO$ content (see Table 1), providing a fast and accurate surrogate for MD simulations throughout the optimization process.

To avoid an exhaustive and inefficient parameter search during the ML optimization procedure, we define a constrained search space and construct an objective function that evaluates how closely the surrogate model's predictions of density and $B^4$ fraction match experimental measurements. Instead of performing a random search, in our approach, candidate parameter sets are evaluated using the trained MLP model, and their outputs are compared to target experimental values via a loss function that quantifies prediction error. This loss serves as the objective for the optimization algorithm, which proposes new candidate parameters accordingly. From each iteration, the top 10 parameter sets, ranked by their predicted performance/prediction error, are selected for full MD simulations (using the glasses from Table 1) to obtain corresponding MD density and $B^4$ fraction values. Subsequently, these new data points are incorporated into the training set, the MLP model is retrained, and the cycle repeats. This iteration allows the surrogate model to progressively improve and concentrate the search on high-potential regions of the parameter space, thereby substantially reducing the number of costly MD evaluations required.

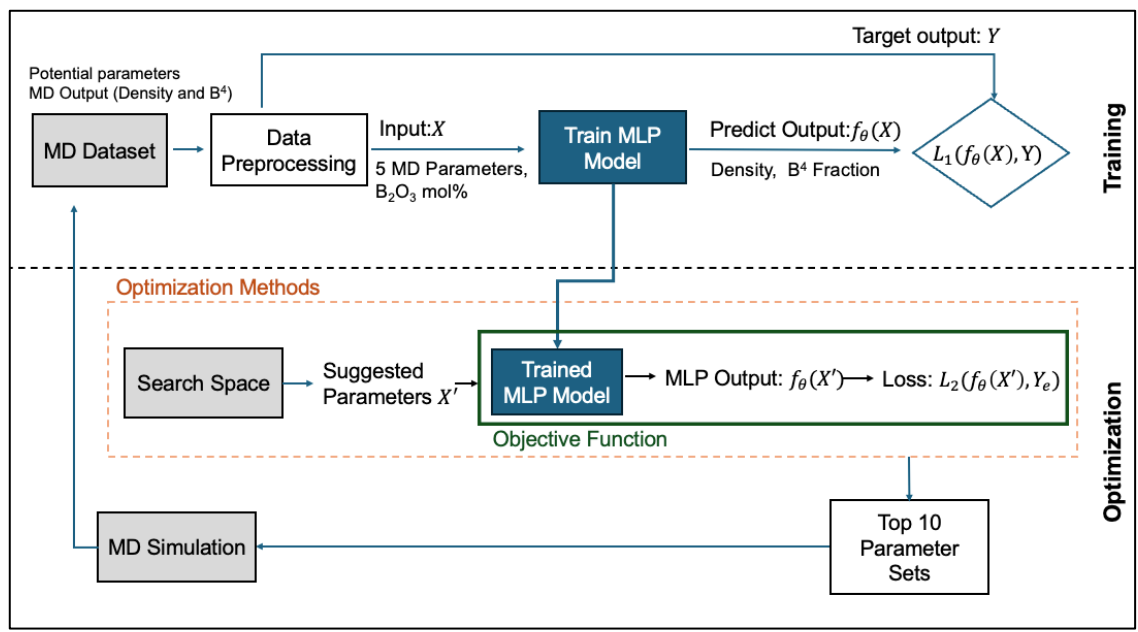

Figure 1: Overview of the active learning framework for potential parameter optimization.

## 2.2 Molecular Dynamics Simulations

Our method is validated using a series of soda-lime borosilicate glasses. We simulated seven glass systems with varying $B_2O_3$ and $SiO_2$ contents, while keeping $Na_2O$ and $CaO$ concentrations approximately fixed near 15 mol% and 10 mol%, respectively. The detailed compositions are provided in Table 1. These compositions were selected because their experimental properties have been previously reported by Smedskjaer et al.[39,40]. Following the naming convention in those earlier studies, each glass is here named according to its nominal (approximate) $B_2O_3$ content.

| Glass ID | Chemical Composition (mol%) | | | |
| --- | --- | --- | --- | --- |
| | $B_2O_3$ | $SiO_2$ | $Na_2O$ | $CaO$ |
| 6B | 4.9 | 69.3 | 16.1 | 9.8 |
| 12B | 10.8 | 63.8 | 14.8 | 10.6 |
| 24B | 21.9 | 51.6 | 15.5 | 11.0 |
| 37B | 38.4 | 36.9 | 14.1 | 10.6 |
| 50B | 49.3 | 24.9 | 15.0 | 10.8 |
| 62B | 62.0 | 12.7 | 14.9 | 10.4 |
| 75B | 74.1 | 0.0 | 15.4 | 10.5 |

Table 1. Compositions of simulated soda-lime borosilicate glasses.

The glasses are simulated using MD simulations with the LAMMPS package. We adopt the classical BKS potential to model the pairwise atomic interactions:

$$U_{ij}(r_{ij}) = \frac{q_i q_j}{r_{ij}} + A_{ij}\, exp\left(\frac{-r_{ij}}{\rho_{ij}}\right) - \frac{C_{ij}}{r_{ij}^6} \qquad (1)$$

where $r_{ij}$ is the distance between atom $i$ and $j$, $q_i$ is the partial charge of atom $i$, and $A_{ij}, \rho_{ij}$ and $C_{ij}$ are empirical parameters specific to the atom pair $(i,j)$. The first term describes the long-range Coulombic interaction, the second term accounts for short-range repulsion, and the third term represents van der Waals attraction. The second and third terms together constitute the Buckingham potential, which is widely used to model short-range interactions in silicate systems. The empirical parameter sets based on the BKS form proposed by Guillot and Sator have demonstrated strong transferability across a wide range of glass compositions[41,42]. In this work, we adopt their parameter values due to this proven performance. However, the Guillot–Sator potential does not include parameters for boron-containing interactions. For the B–Si pair, we employ values from Kieu's potential[16]. Our primary objective is to optimize the parameter values for B–O and B–B interactions based on available experimental data. For the B-B pair, we set the $C$ term to zero by design, as boron atoms in silicate glasses are typically connected via bridging oxygens rather than direct B–B bonds. Thus, the total number of parameters to be optimized is five: $A_{B-O}, \rho_{B-O}, C_{B-O}, A_{B-B}$, and $\rho_{B-B}$.

MD simulations were carried out using a standard melt–quench–relax protocol[43]. Interatomic interactions were modeled using a Buckingham and Coulombic potential with an 11 Å cutoff for both short-range and electrostatic terms. Long-range electrostatics were treated using the particle–particle particle–mesh (PPPM) method with a precision of $1.0\times10^{-5}$. Partial charges for each atom type were taken from the original Guillot–Sator interatomic potential, as detailed in the Supplementary Information, S1[38]. The five unknown B–O and B–B interaction parameters were initially sampled using a combination of grid search and random search to generate a diverse set of trial parameter combinations. Each parameter set was used to simulate each of the seven glass compositions individually.

For each composition, approximately 3000 atoms were randomly placed within a cubic simulation cell while avoiding unphysical overlaps. A timestep of 1.0 fs was used throughout the simulations. The initial configuration was first energy-minimized and then equilibrated under the NVT ensemble at 300 K for 20 ps to relax any residual stresses. This was followed by a 20 ps NPT equilibration at 300 K and 40,000 atm to compact the structure. To ensure proper melting, the system was heated to 3000 K and equilibrated for 100 ps under NPT conditions at 20,000 atm. This high-pressure phase helps maintain atomic cohesion at elevated temperatures. The pressure was then released to 0 atm, and the system was further equilibrated at 3000 K for another 100 ps, producing a well-equilibrated liquid state. The glass formation process was carried out by cooling the liquid from 3000 to 300 K under ambient pressure over 2.7 ns at a cooling rate of 1 K/ps, allowing for gradual structural freezing. After cooling, the system was further equilibrated under NPT conditions at 300 K and 0 atm for 200 ps to ensure structural relaxation.

To characterize the properties of the resulting glass, we performed an additional 100 ps NVT simulation at 300 K, during which atomic configurations were recorded every 1 ps, resulting in 100 evenly spaced snapshots. These configurations were used to calculate the average glass density and fraction of $B^4$, providing statistically reliable structural and thermodynamic properties of the glassy state. The density was computed by averaging the instantaneous density values over the final 100 ps of the NVT simulation. The $B^4$ fraction, defined as the ratio of fourfold-coordinated boron atoms to the total number of boron atoms, was determined through coordination analysis. A boron atom was considered fourfold coordinated if it had four oxygen neighbors within a 1.8 Å cutoff, a value determined based on the radial distribution function (RDF)[36]. These two structural properties, density and $B^4$ fraction, served as target metrics for training and evaluating the surrogate MLP model.

## 2.3 Training

In the initial MD dataset, each data point consists of a five-dimensional parameter vector and its corresponding MD-computed density and B⁴ fraction across all seven glass compositions listed in Table 1. To stabilize model training, both the input parameters and the density values were normalized using min–max scaling to the range [0,1], as they originally spanned different numerical ranges. In contrast, the B⁴ fraction naturally lies within the range [0,1] and thus required no further scaling. This normalization ensures that all input features and target outputs are on comparable scales during training. However, it is noted that the scaling ranges are not completely equivalent, which is why we will introduce a weighting factor during optimization later.

To prepare the data for training, each input sample is constructed by concatenating the five potential parameters with the $B_2O_3$ molar fraction of a specific glass composition. The corresponding outputs are the density and B⁴ fraction predicted by MD for that composition. As a result, the MLP takes six input features and produces two outputs. The model is trained by minimizing the sum of mean squared errors (MSE) between the predicted and MD-computed density and B⁴ values. The loss function is defined as,

$$L = \frac{1}{N}\sum_{i=1}^{N}[(\hat{d}_i - d_i)^2 + (\hat{B}_i^4 - B_i^4)^2] \quad (2)$$

where $\hat{d}_i$ and $\hat{B}_i^4$ are predicted density and B⁴ fraction, respectively, for the $i$th data point, and $d_i$ and $B_i^4$ are the corresponding MD-computed values. $N$ is the number of data points in the batch (or dataset) used during training or evaluation. Because the input features and output targets are normalized to comparable scales, this unweighted MSE loss treats density and B⁴ fraction with balanced importance during training. This loss formulation enables the MLP to function as a surrogate model for MD simulations, offering fast and reasonably accurate predictions of glass properties.

By explicitly including the $B_2O_3$ composition as an input, the model becomes composition aware. It can generalize across different boron concentrations in the selected glass system, rather than requiring separate models for each composition. This design mimics the MD simulation setup, where potential parameters and composition jointly determine the resulting structure, and allows the MLP to efficiently guide the optimization process. As more simulation data are added iteratively, the surrogate model continues to improve in accuracy, further enhancing its ability to identify promising parameter sets.

**2.4 Optimization**

Building on the previously described framework, the optimization process is designed to be adaptive and iterative. It follows an active learning approach, where insights from previous steps are used to refine the parameter space, update the objective function, and improve the surrogate model. Each iteration in this framework consists of three main steps: training the surrogate MLP model using the updated MD dataset, optimizing candidate parameter sets using a surrogate-guided algorithm, and evaluating selected candidates through MD simulations. The resulting MD-computed properties are then added back into the MD dataset to support the next iteration. This dynamic feedback loop allows the framework to concentrate computational resources on the most promising regions of the parameter space and accelerate convergence toward optimal parameters. Specific choices for the search range and loss formulation are discussed in the Results section. The following optimization algorithms were employed during the parameter optimization process.

### 2.4.1 Bayesian optimization

As an initial approach to parameter optimization, we employed Bayesian Optimization, or BayOpt, to efficiently explore the five-dimensional parameter space[44]. BayOpt constructs a probabilistic surrogate model of the objective function, typically using a Gaussian process, to estimate both the expected value and uncertainty of the function at untested points. An acquisition function is then used to select the next candidate by balancing exploration, which samples uncertain regions, and exploitation, which focuses on promising areas.

The objective function takes a candidate parameter set as input. For each glass composition, the five-dimensional parameter vector was concatenated with the $B_2O_3$ composition and passed into the trained MLP model to predict the density and $B^4$ fraction. The MSE was then computed between the predicted and experimental values for each property and composition. The final scalar output of the objective function, denoted as $L_{BayOpt}$, is defined as the sum of the MSE for normalized density and $B^4$ fraction across all compositions,

$$L_{BayOpt} = MSE_d + MSE_B \qquad (3)$$

where

$$MSE_d = \sum_{i=1}^{7}(\hat{d}_i - d_i^e)^2 \qquad (4)$$
$$MSE_B = \sum_{i=1}^{7}(\hat{B}_i^4 - B_i^{4e})^2 \qquad (5)$$

Here, $\hat{d}_i$ and $\hat{B}_i^4$ are predicted density and $B^4$ fraction, respectively, from the trained MLP model, while $d_i^e$ and $B_i^{4e}$ are the corresponding experimental values. This composite loss guided the optimization toward parameter sets that simultaneously match both target properties.

In each iteration, we executed BayOpt five independent times to reduce the risk of convergence to local minima. To further balance exploration and exploitation, hyperparameters such as the acquisition function type and the number of initial sampling points were varied across runs based on performance in previous iterations. Each BayOpt run returned the top five candidate parameter sets, ranked by their objective function values, which reflect the predicted deviation between model outputs and experimental targets. From this pool of 25 candidates, we selected the overall top 10 parameter sets to be evaluated using full MD simulations. To determine the best-performing set in each iteration, we computed the MSE in the same way as defined in the objective function, but this time comparing MD simulation outputs to experimental targets. The parameter set with the lowest total MSE was selected as the final result for that iteration. The resulting MD-computed properties were then added to the training dataset, enabling the surrogate model to improve in subsequent iterations.

### 2.4.2 Covariance Matrix Adaptation Evolution Strategy (CMA-ES)

After several outer iterations using BayOpt, we also employed the Covariance Matrix Adaptation Evolution Strategy, or CMA-ES, as an alternative optimization method[45].CMA-ES is a derivative-free evolutionary algorithm inspired by natural selection principles, and it does not require an explicit surrogate model or acquisition function. Within each outer iteration, where the surrogate MLP model is retrained using newly acquired MD data, CMA-ES is used as the internal optimizer to generate new candidate parameter sets. The CMA-ES algorithm begins with an initial estimate of the parameter mean and covariance matrix. During its running, it samples a population of candidate solutions from a multivariate normal distribution. The mean represents the current best estimate of the optimal parameters, while the covariance matrix captures the shape and orientation of the search space. These candidates are evaluated using the trained MLP model, and the top-performing solutions are used to update both the mean and the covariance matrix.

This adaptive mechanism biases future sampling toward more promising regions and allows the algorithm to refine its search direction and scale over time. To ensure robustness and diversity in sampling, we followed the same strategy as used for BayOpt, that is, CMA-ES was executed five times per iteration, each run producing five top-ranked candidate parameter sets. From this pool of 25 candidates, the overall top 10 was selected and evaluated via full MD simulations. CMA-ES is particularly effective for high-dimensional and noisy optimization problems, and its robustness and flexibility make it well-suited for complex, black-box objective landscapes where gradient information is unavailable.

## 3. Results

### 3.1 MD Dataset

The MD dataset is collected through MD simulations as described in Section 2.2. To systematically explore the BKS potential parameter space, we adopted the optimized BKS parameter value from Wang et al. previous study as our starting point, referred to here as the original parameter set[36]. Using this reference as a baseline, we generated 230 distinct parameters sets using a combination of grid search and random sampling within a defined percentage. Each of these parameter sets was then used as the BKS potential parameters in MD simulations to compute the corresponding density and B⁴ fraction. The resulting parameter–property pairs form our initial MD dataset. This hybrid sampling strategy was designed to ensure comprehensive coverage of the parameter space. The sampling range was set to ±15% relative to the original values. One exception was made for the $\rho_{B-O}$ parameter, as preliminary simulations revealed that reducing this parameter by more than 5% often caused instability in atomic configurations, such as atom overlap and unphysical bonding. To mitigate such issues, the sampling range for $\rho_{B-O}$ was limited to [–5%, +15%], while the full [–15%, +15%] range was retained for the other parameters.

Figure 2 shows the distribution of each of the five parameters across all 230 sampled sets. Each parameter set was evaluated for seven distinct borosilicate glass compositions, resulting in a total of 1,610 data points used to train and test the MLP model. Of these, 70% were allocated for training and the remaining 30% for testing. The surrogate model's hyperparameters were adjusted as needed during each iteration to maintain or improve performance. The model employed a ReLU activation function and was optimized using the Adam optimizer, with L2 regularization applied via weight decay to reduce overfitting. A learning rate scheduler was also used to automatically lower the learning rate when performance plateaued, helping to ensure stable convergence.

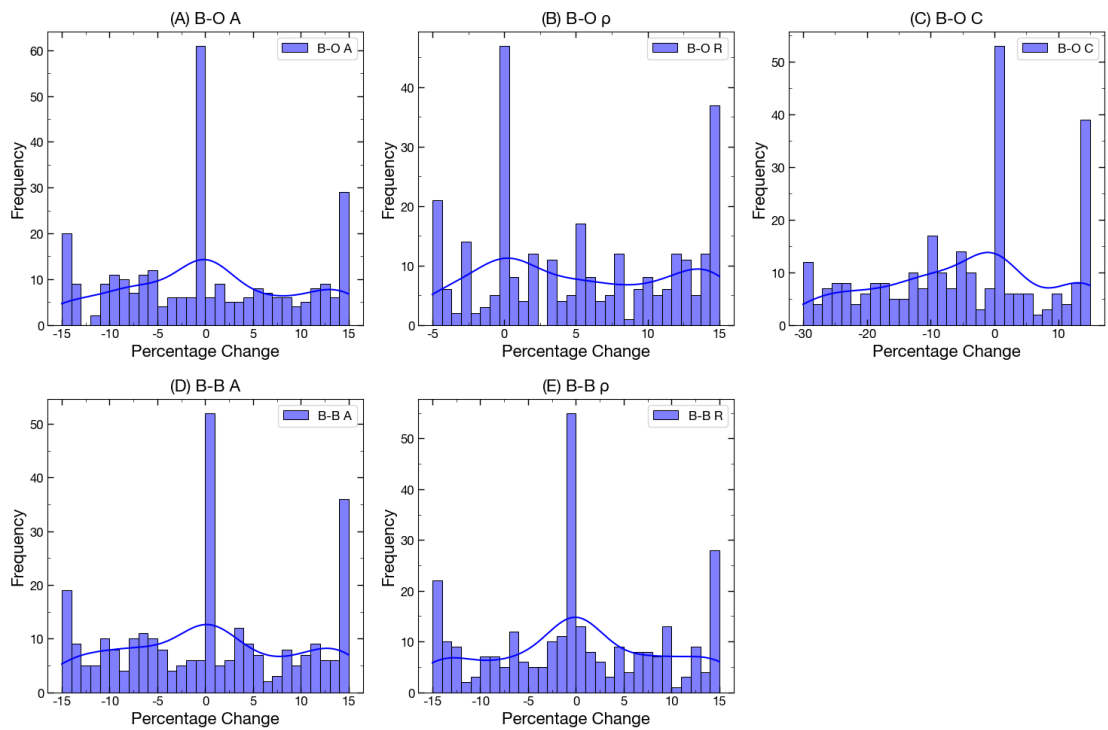

Figure 2. Distributions of percentage changes for the five potential parameters related to the original parameter set value across all 230 sampled parameter sets. Each subplot corresponds to a different parameter: (A) $A_{B-O}$, (B) $\rho_{B-O}$, (C) $C_{B-O}$, (D) $A_{B-B}$, and (E) $\rho_{B-B}$.

## 3.2 Bayesian Optimization

At the start of our process, we used Bayesian Optimization (BayOpt) as the first optimization algorithm. At the beginning of the optimization, the appropriate search range for parameter variation was unknown. To balance the risk of exploring a space that was too broad, which could lead to instability, or too narrow, which could limit discovery, we did not use the same broad range as that used to generate the initial MD dataset in Section 3.1. While the MD dataset was designed to cover a wider parameter space to train a robust surrogate model, the optimization range was intentionally narrowed to ensure numerical stability and facilitate efficient convergence. Therefore, we selected a moderate initial range of ±10% relative to the original parameter values for all five interatomic potential parameters. One exception was made for the $\rho_{B-O}$ parameter, which had already been shown to become unstable if decreased by more than 5%. Therefore, we constrained the $\rho_{B-O}$ range to [–5%, +10%].

Figure 3A presents the best parameter set from each iteration of the BayOpt process. For interpretability, we report the root mean squared error (RMSE) instead of MSE in the figure. The RMSE is calculated based on the trained MLP model predicting density and $B^4$ values compared with the experimental target values. As described in Section 2.3, the MLP model was trained on a normalized MD dataset, where density values were min–max normalized while $B^4$ values remained unnormalized because they already fall within the [0, 1] range. Therefore, the MLP model outputs used here reflect normalized density and unscaled $B^4$ values, and we refer to them as such throughout the results. The RMSE is shown separately for normalized density and $B^4$ fraction, and the total RMSE is defined as the sum of both components. Iteration 0 corresponds to the original parameter set and serves as the baseline. We find a clear downward trend in density RMSE, while $B^4$ RMSE remains relatively stable compared to the baseline. This indicates that BayOpt was effective in improving the fit to density targets but had limited success in reducing the error in $B^4$ fraction. The best-performing parameter set was obtained in iteration 6, which achieved both the lowest density RMSE and the lowest total RMSE. We refer to this parameter set as the BayOpt result set in the following sections.

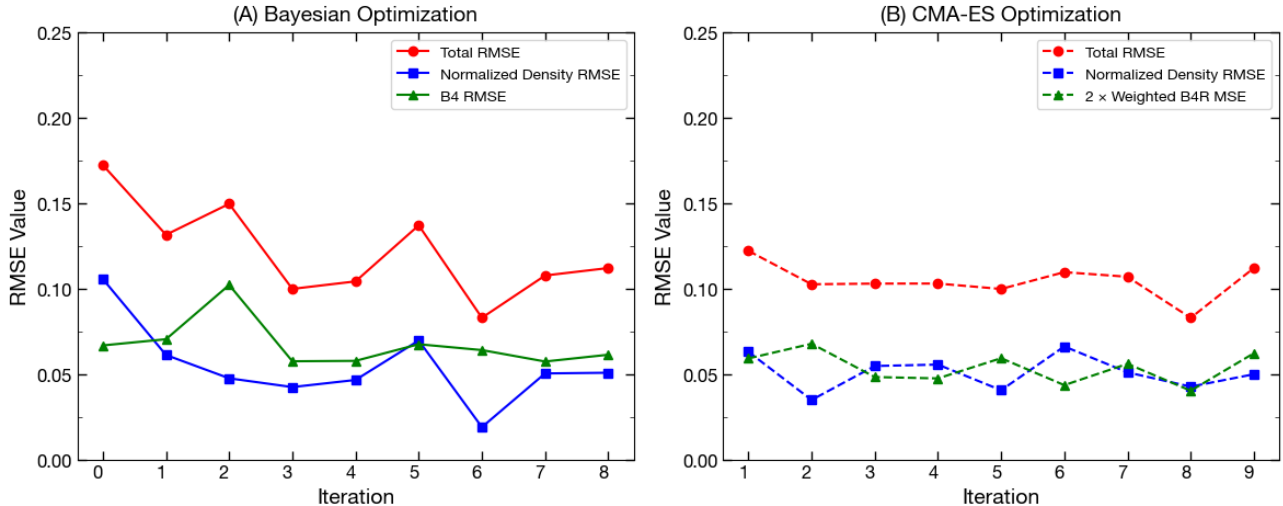

Figure 3. RMSE trends in normalized density, $B^4$ fraction, and their sum for the best parameter set of each iteration: (A) BayOpt and (B) CMA-ES.

Details of the best BayOpt result set for each glass composition are shown in Figures 4A-B, compared against experimental targets. For each glass composition, MD simulations were performed using the optimized parameter set with six different initial configurations to estimate uncertainty and ensure result stability. The error bars shown in Figure 4 represent the mean ± one standard deviation of these six MD runs, capturing the variability due to different initial atomic configurations. Experimental targets are shown with error bars derived from the original experimental sources. Figure 4A compares simulated and experimental density values across varying $B_2O_3$ compositions. The optimized results closely match the experimental values. Figure 4B shows the $B^4$ fraction results. While the simulated $B^4$ values agree well with experiments at lower $B_2O_3$ concentrations, accuracy declines when the $B_2O_3$ content exceeds 40%. This observation aligns with previous findings that BayOpt performs well in improving density predictions but is less effective at optimizing $B^4$ fraction, especially at higher $B_2O_3$ concentrations.

Additionally, as the optimization progressed, we identified key limitations of BayOpt in the context of our problem. The primary challenge stems from the high dimensionality of the parameter space, where five interdependent potential parameters must be optimized simultaneously. In such settings, the number of samples required to adequately explore the space increases exponentially, resulting in severe data sparsity. The surrogate probabilistic model, typically a Gaussian process, struggles to accurately estimate the objective function, especially in unexplored or sparsely sampled regions. This high level of uncertainty makes it difficult for the acquisition function to reliably guide the search toward truly promising areas. As a result, the optimization may either over-explore uncertain regions without significant improvement or over-exploit already sampled areas, leading to premature convergence or stagnation. Furthermore, the computational cost of BayOpt becomes a bottleneck as the dataset grows. Training the surrogate model becomes slower, and evaluating and optimizing the acquisition function in a high-dimensional space becomes increasingly resource-intensive. These limitations make BayOpt less efficient and less scalable for our problem, where parameter interactions are complex and function evaluations via MD simulations are computationally expensive.

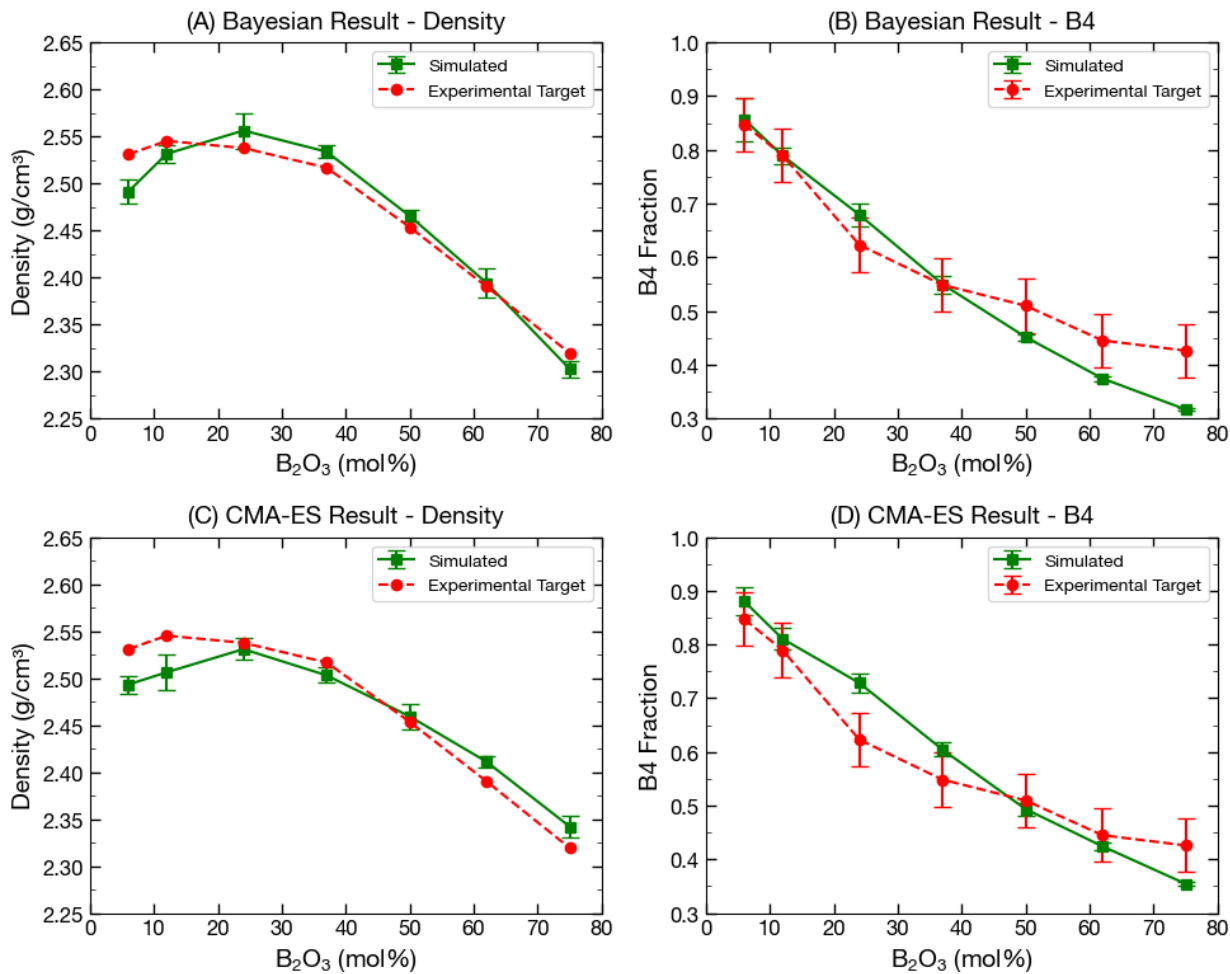

Figure 4. Comparison of simulated and experimental results for the best parameter sets.
(A) Density and (B) B⁴ fraction results from the best BayOpt parameter set.
(C) Density and (D) B⁴ fraction results from the best CMA-ES parameter set.

## 3.3 Search Range Update

After completing multiple iterations of BayOpt, we sought to evaluate whether the initial search range was sufficient. To do this, we analyzed how the distribution of the top 10 selected parameter sets evolved over iterations. We focused on the three B–O interaction parameters ($A$, $\rho$, and $C$), which had the most significant influence on optimization performance. The results are shown in Figure 5. Figure 5A displays the distribution of $A_{B-O}$ values across iterations, while Figures 5B and 5C show the corresponding distributions for $\rho_{B-O}$ and $C_{B-O}$, respectively. In each plot, the horizontal axis represents the iteration number, and the vertical axis indicates the percentage change relative to the original parameter value. Each black dot represents one of the top 10 selected values for that iteration. Each rectangular bin reflects the number of points falling within a specific percentage range. The color bar beside each plot shows how the background shading corresponds to the number of points in each bin, with darker colors indicating a higher concentration of points in that region. A red dashed vertical line separates the final four BayOpt iterations result using initial search range (left side) from later iterations using an expanded search space (right side). To the left of the red line, the search space was initially set to [–10%, +10%] for all parameters, except $\rho_{B-O}$, which was constrained to [–5%, +10%] to prevent simulation instability. However, the optimization results revealed a clear boundary effect, that is, selected parameter values for $A_{B-O}$ and $C_{B-O}$ frequently clustered near the boundaries of the allowed range. This consistent boundary behavior suggested that higher-performing solutions might exist outside the initial range, motivating the need to expand the search space in subsequent iterations.

To address the observed boundary clustering, we expanded the search space to [–15%, +15%] for most parameters, while maintaining $\rho_{B-O}$ within [–5%, +15%] to ensure simulation stability. However, this wider parameter space further highlighted the limitations BayOpt discussed earlier. As the search space expanded, the Gaussian process surrogate model faced greater difficulty in accurately modeling the objective function, especially in sparsely sampled regions. The increased uncertainty made it harder for the acquisition function to effectively guide the search, leading to slower convergence and less efficient exploration. To overcome these challenges, in the following we transition to CMA-ES, which is inherently well-suited for navigating high-dimensional and wide parameter spaces[46]. Its robustness and flexibility make it a better fit for our problem, where parameter interactions are complex and the objective landscape is difficult to model using probabilistic surrogates.

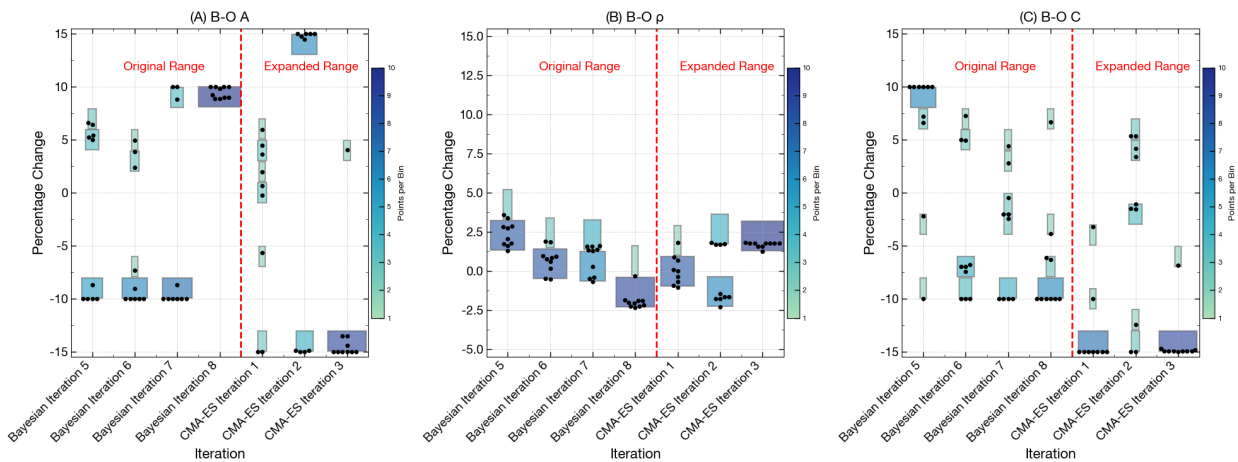

Figure 5. Distributions of top 10 selected parameter values for B-O pair across optimization iterations. The horizontal axis shows the iteration number, consistent with the iteration numbering used in Figure 3.

The right side of the red line in Figure 5 presents the results from the first three iterations following the transition to CMA-ES. For the $A_{B-O}$ parameter, the selected values shifted markedly into the expanded region, with a notable concentration near both the new upper and lower bounds, which were previously inaccessible using BayOpt. A similar trend was observed for $C_{B-O}$, where most selected values clustered near the lower boundary of the extended range. In contrast, the distribution of $\rho_{B-O}$ values remained relatively unchanged, consistently falling within a narrow band between approximately –2.5% and +5%. This suggests that the original search range for $\rho_{B-O}$ was already sufficient, and further expansion did not yield significantly different selections. A more detailed analysis of parameter distributions during optimization can be found in Supplementary Information Section S2.

These findings confirm that the initial search space was overly restrictive for $A_{B-O}$ and $C_{B-O}$, and that expanding it enabled the identification of improved parameter sets. However, the continued boundary clustering for both parameters indicates that even the expanded range may remain insufficient and could benefit from further extension. Overall, this analysis underscores the effectiveness of the active learning framework in dynamically refining the optimization strategy based on intermediate results. By enabling rapid identification of search space limitations and redirecting exploration toward more promising regions, the framework reduces the number of costly MD simulations required. This adaptability is particularly advantageous when optimizing in high-dimensional, complex parameter landscapes.

## 3.4 CMA-ES

As discussed previously, after completing eight iterations using BayOpt, we transitioned to CMA-ES as our optimization method. CMA-ES iterations began with the expanded search ranges. However, by iteration 6, we observed that the selected values for $A_{B-O}$ and $C_{B-O}$ continued to cluster at the boundaries of the expanded ranges. This indicated that further improvement was possible and that the current limits were still too restrictive for these two parameters. To address this, we further adjusted their search windows. For $A_{B-O}$, which consistently approached both the upper and lower boundaries, we extended the range to [–20%, +20%]. In contrast, $C_{B-O}$ values repeatedly pushed only the lower bound. To avoid unnecessary expansion on the upper end, we applied a directional adjustment by shifting to a non-symmetric window of [–25%, +5%]. The search ranges for the remaining parameters were left unchanged, as their selected values consistently fell within the existing limits and showed no signs of boundary saturation.

Since BayOpt was less effective at improving the $B^4$ fraction particularly in high-$B_2O_3$ compositions, we revised the objective function to better prioritize these cases. To enhance the accuracy of $B^4$ predictions where earlier iterations had underperformed, we introduce a composition-weighted loss formulation. Specifically, instead of treating each glass composition equally, we applied a weighting scheme to the $B^4$ MSE term based on the relative $B_2O_3$ content of each composition. For each composition $i$, the weight $w_i$ was defined as,

$$w_i = [B_2O_3]_i / \sum_{j=1}^{7}[B_2O_3]_j \tag{6}$$

The total weighted $B^4$ MSE was computed as,

$$MSE_{B\_weighted} = \sum_{i=1}^{7} w_i (\hat{B}_i^4 - B_i^{4e})^2 \tag{7}$$

where $\hat{B}_i^4$ is the trained MLP predicted $B^4$ value for composition $i$ and $B_i^{4e}$ is the experimental $B^4$ value. This weighting approach emphasizes compositions with higher $B_2O_3$ content during optimization. The final objective function was then defined as a weighted sum of the density MSE and the weighted $B^4$ MSE,

$$L_{CMA-ES} = MSE_d + w_B MSE_{B\_weighted} \tag{8}$$

where $w_B = 2$ is a scalar that increases the influence of $B^4$ accuracy on the total loss. The definition of $MSE_d$ is the same as in Eq. (4).

The selection of the best parameter set in each CMA-ES iteration was based on the same loss function used during optimization, with the key difference that the loss was computed using MD simulation results instead of the surrogate model predictions. The iteration-wise performance of CMA-ES is illustrated in Figure 3B, showing the RMSE for normalized density and weighted $B^4$ RMSE. The latter was calculated using a composition-weighted formulation for each glass composition and scaled by a factor of two, in accordance with the objective function definition. The total RMSE is reported as the sum of these two components. The results illustrate a trade-off between density and $B^4$ accuracy, where improvement in one metric slightly compromises the other. Notably, after the new expanded search ranges were introduced in iteration 6, both metrics showed marked improvement. By iteration 8, CMA-ES achieved the most balanced performance, i.e., both the normalized density RMSE and the weighted $B^4$ RMSE reached their lowest values. This indicates that the updated search space enabled the identification of significantly better-performing parameter sets. Accordingly, the parameter set obtained in iteration 8 was selected as the final result for CMA-ES.

Figures 4C-D present the detailed MD simulation results of the best-performing CMA-ES parameter set, compared with the experimental targets. The density predictions align well with the experimental values, consistent with the results obtained using BayOpt. However, differences emerge in the $B^4$ predictions. The CMA-ES result shows improved agreement with experimental $B^4$ values at higher $B_2O_3$ concentrations (above 50 mol%), as intended, but exhibits reduced accuracy at lower $B_2O_3$ levels. This trend contrasts with the BayOpt result, which performed better

at low-$B_2O_3$ concentrations but failed to capture the high-boron regime accurately. These results demonstrate the effectiveness of our framework in identifying optimal parameter sets. Using active learning and dynamic search space refinement, our method successfully converged to a high-performing solution in fewer than ten iterations, significantly reducing the number of costly MD simulations. Moreover, the framework allowed us to prioritize specific properties, such as improving $B^4$ accuracy in high-$B_2O_3$ glasses, by modifying the objective function. The reduced agreement at lower $B_2O_3$ concentrations may reflect an inherent trade-off in simultaneously optimizing $B^4$ fraction accuracy across a wide range of $B_2O_3$ compositions, which will be further discussed in Section 4.

### 3.4.1 Structure Analysis

We now assess the structural accuracy of simulated glasses using optimized parameters. Beyond density and $B^4$ predictions, more comprehensive structural analyses—including partial pair distributions, bond angle distributions, ring size distributions, and more. Detailed analysis can be found in ref. [38]. Here, we focus specifically on comparing simulated X-ray structure factors $S(Q)$ with experimental measurements, taken from ref. [38], for two representative glass compositions, 6B and 62B, as shown in Figure 6. These two systems serve as examples of low- and high-boron glasses, respectively. Full comparisons across all seven compositions are provided in the Supplementary Information, S3.

Despite producing different densities and $B^4$ fractions, the two potential parameter sets produce similar X-ray structure factors, particularly in the low-$Q$ region such as the first sharp diffraction peak (FSDP) at around 2 Å$^{-1}$. As the FSDP is typically assigned to ring structures in silicate glasses[47,48], this similarity suggests that medium-range structural ordering remains largely unaffected by the parameter differences. However, notable discrepancies emerge in the high-$Q$ region between the two parameter sets. These differences demonstrate that parameter modifications primarily influence short-range atomic correlations, with the effect becoming more pronounced at higher boron concentrations. This trend is expected since the potential energy terms directly govern atomic bonding interactions, and these terms are fundamentally controlled by the empirical parameters.

Our optimized parameter sets demonstrate strong agreement with experimental X-ray structure factor measurements. FSDP is reproduced across most compositions, particularly for low-boron glasses, indicating that our simulations successfully capture the medium-range structural organization characteristic of these glasses. For high-boron compositions, some discrepancies emerge between simulated and experimental results. These deviations align with our earlier observations regarding $B^4$ prediction accuracy in boron-rich systems. We hypothesize that the discrepancies in high-boron compositions may relate to fictive temperature differences between experimental synthesis conditions and MD simulation. While our data indicate that the relative $B^4$ fraction is actually less sensitive to fictive temperature variations in boron-rich systems, the absolute magnitude of boron coordination changes can still be significant due to the higher total $B_2O_3$ content, potentially contributing to the observed deviations in structural predictions.

Given that our optimization process focused exclusively on matching glass densities and $B^4$ fractions, the ability of our simulated $S(Q)$ to reproduce the major structural peaks and overall trends represents an encouraging validation. The structural fidelity achieved suggests that our parameter sets capture the essential physics governing these glass systems, even though some compositional limits remain for highly accurate quantitative predictions. This performance demonstrates that targeted parameter optimization, even with limited experimental constraints, can yield models capable of reproducing key structural characteristics across a range of glass compositions.

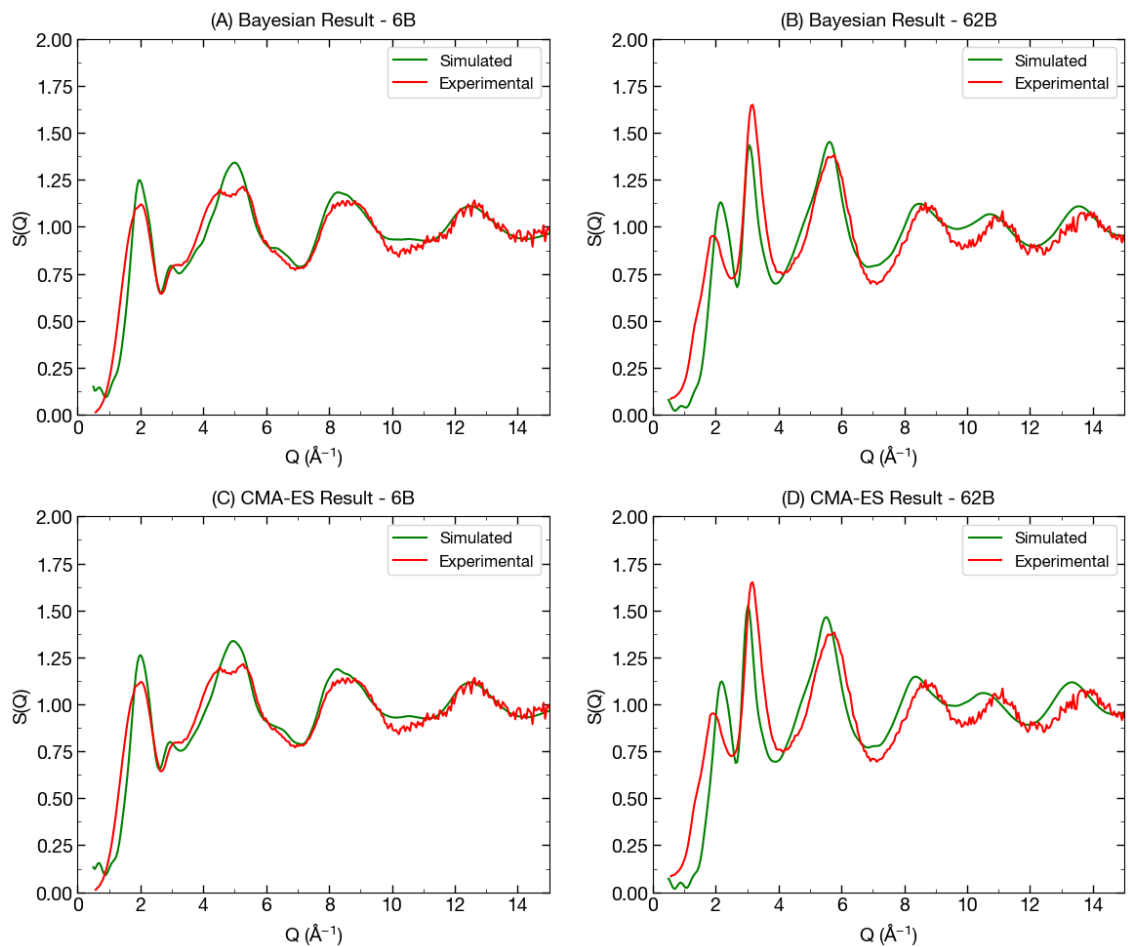

Figure 6. Comparison of simulated and experimental structure factors (from ref. [38]) for the best parameter sets. (A-B) Results from the best BayOpt parameter set for (A) 6B and (B) 62B. (C-D) Results from the best CMA-ES parameter set for (A) 6B and (B) 62B.

## 4. Discussion

### 4.1 Trade-off Analysis

#### 4.1.1 Density vs B⁴ trade-off

During the optimization process, we identified a fundamental trade-off between density and B⁴ fraction accuracy. As shown in the CMA-ES results (Figure 3B), due to the updated objective function adding more weight on B⁴ MSE, efforts to improve B⁴ predictions often resulted in increased error in density predictions, and vice versa. In the early CMA-ES iterations, for example, reductions in B⁴ RMSE were frequently accompanied by higher density RMSE. This pattern underscores an intrinsic tension between the two objectives and suggests that simultaneously minimizing both errors is nontrivial. This trade-off is further illustrated by a specific parameter set from the initial MD dataset, which was selected because it achieved the lowest B⁴ MSE among all dataset. As shown in Figure 7, this set produced the most accurate B⁴ predictions across all compositions in the dataset. However, its corresponding density values deviated significantly from experimental measurements, demonstrating the challenge of achieving simultaneous optimization of both structural properties. More detail about the density and B⁴ trade-off can be find Supplementary Information, S4.

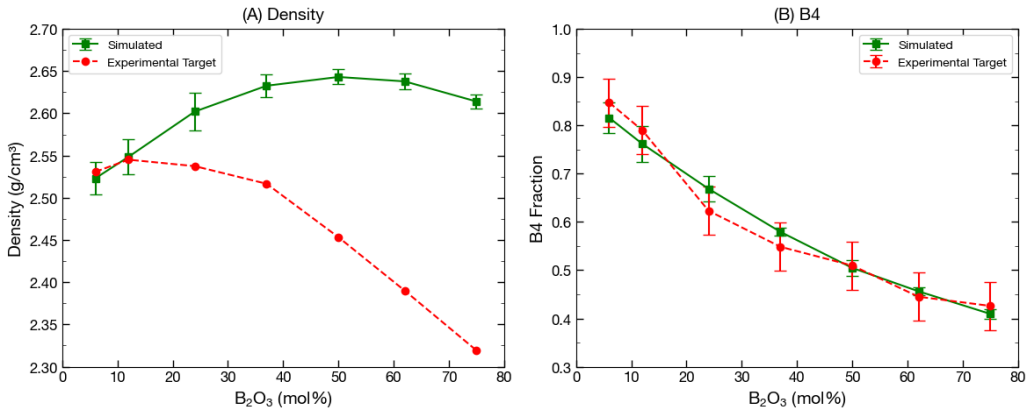

Figure 7. Simulated results of a parameter set with high B⁴ accuracy but poor density agreement.

To further investigate the origin of the observed trade-off and gain deeper insight into the optimization landscape, we conducted a parameter sensitivity analysis focusing on the $A_{B-O}$ and $C_{B-O}$ parameters. These two parameters were selected due to their strong influence on the results and the fact that their search ranges were revised twice during the optimization process. The other three parameters were held constant (i.e., as the original values), while 150 values were uniformly sampled for each $A_{B-O}$ and $C_{B-O}$ parameter within a ±15% range relative to their baseline values. This resulted in a grid of 150×150 parameter combinations, each of which was evaluated using the surrogate MLP model trained at the 5th CMA-ES iteration. Then, the normalized density and fraction of B⁴ MSEs computed according to Eqs. (4) and (5), and the total MSE was calculated as their sum, following Eq. (3).

The resulting error landscapes are presented in Figure 9, where lighter colors indicate lower error. In panel (B), which represents the MSE for normalized density, the optimal region is located in the lower portion of the space, suggesting that smaller $C_{B-O}$ values consistently lead to lower density errors. In contrast, panel (C), which shows the MSE for B⁴ fraction, exhibits a far more complex contour. This indicates that $A_{B-O}$ and $C_{B-O}$ interact in a nontrivial manner in shaping the B⁴ accuracy, with no clear monotonic trend across the parameter space. This observation helps explain why the optimization initially succeeded in reducing density error using BayOpt but struggled to improve B⁴ accuracy. The relatively smooth landscape for density MSE enabled more straightforward convergence, while the more intricate and irregular landscape for B⁴ MSE posed a greater challenge. The divergence between the optimal regions for density and B⁴ further confirms that the directions for minimizing these two objectives are not aligned within the parameter space. Panel (A), which presents the total MSE, reveals a highly nonlinear and irregular optimal region. Even within this two-dimensional slice of the full five-dimensional space, the optimization landscape is complex, highlighting the intrinsic difficulty of locating a global optimum through direct search or standard surrogate-guided methods.

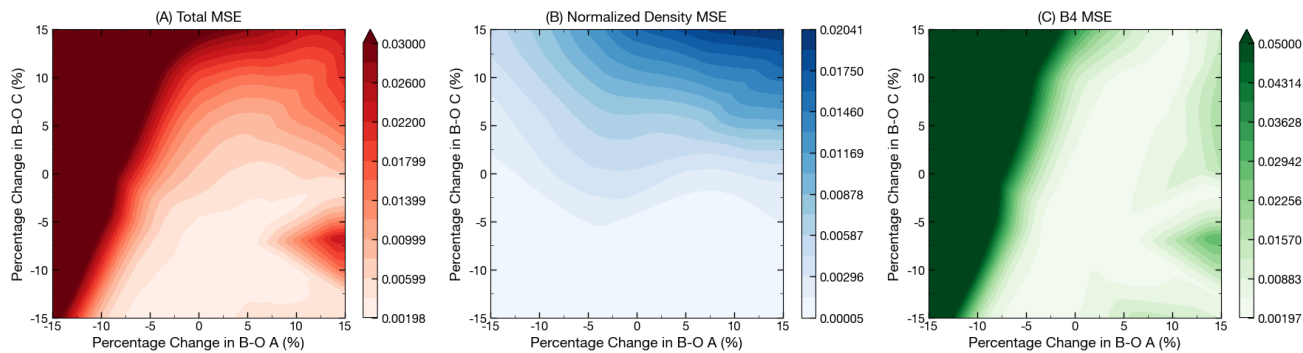

Figure 9. Parameter sensitivity analysis for $A_{B-O}$ and $C_{B-O}$. (A) Total MSE surface, (B) normalized density MSE, and (C) B⁴ MSE.

In summary, our results reveal a fundamental trade-off between optimizing density and B⁴ fraction in borosilicate glasses. These two properties are structurally coupled but governed by distinct mechanisms and respond differently to variations in interatomic potential parameters. Density reflects the global compactness of the atomic network and is influenced by both atomic packing efficiency and the molar mass of constituent elements. In contrast, the B⁴ fraction characterizes the local coordination environment of boron atoms and is directly related to the connectivity of the glass network via the formation of bridging oxygens. The coordination state of boron is highly sensitive to glass composition and thermal history, making it particularly challenging to capture accurately using classical interatomic potentials.

This trade-off is especially pronounced in BKS-style fixed-charge, pairwise interaction models, which lack the flexibility to decouple local bonding behavior from global structure. For example, increasing the B–O bond strength or modifying partial charges to favor $B^4$ formation tends to increase network rigidity and compaction, often resulting in overestimated densities. Conversely, tuning the potential to reproduce experimental density may require weakening B–O interactions or adjusting charges in a way that under-stabilizes tetrahedral boron, leading to an underprediction of $B^4$ fraction. Because fixed-charge models do not support charge transfer or coordination-adaptive behavior, they are inherently limited in their ability to independently optimize both properties. This limitation has been noted in the literature, where empirical potentials calibrated to reproduce $B^4$ often overpredict glass stiffness or density, and more transferable models must compromise between these competing structural targets[9,49,15,47].

### 4.1.1 Trade-off between high-boron and low-boron compositions

Another key trade-off identified during the optimization process lies between accurately modeling low- and high-boron glass compositions. As shown in the $B^4$ results from BayOpt (Figure 4B), the model performs well at lower $B_2O_3$ concentrations but underperforms at higher concentrations. In contrast, the CMA-ES results (Figure 4D) show improved $B^4$ accuracy at high $B_2O_3$ compositions, following the intentional reweighting of the objective function to prioritize performance in this region. However, this improvement comes at the cost of reduced accuracy in low-boron compositions, illustrating a composition-dependent trade-off.

This challenge again arises from the variable coordination behavior of boron atoms. In our simulations, we use a fixed-charge BKS-type potential where boron and oxygen atoms are assigned constant partial charges. However, in reality, a $BO_4$ unit carries an additional negative charge compared to a $BO_3$ unit, a distinction that is not captured in fixed-charge models. As a result, tuning the B–O interaction parameters and charges to reproduce the correct $B^4$ fraction in high-$B_2O_3$ (low-$SiO_2$) glasses can lead to underestimation of $B^4$ in low-$B_2O_3$ (high-$SiO_2$) compositions. This is because although the modifier content (e.g., $Na^+$, $Ca^{2+}$) remains fixed across the studied compositions, low-$B_2O_3$ glasses have fewer boron atoms, so each boron has access to more modifier cations. This makes it easier to form $BO_4^-$ units through charge compensation. On the other hand, high-$B_2O_3$ glasses contain more boron atoms, meaning the same number of modifiers must be distributed among more boron units, reducing the availability of modifier cations per boron and makes it harder to stabilize $BO_4^-$ units. As a result, if the potential is optimized for low-$B_2O_3$ glasses, it tends to overestimate the $B^4$ fraction in high-$B_2O_3$ composition.

In summary, the two key trade-offs identified in our simulations, between fitting density and B⁴ fraction and between achieving accuracy at high and low $B_2O_3$ compositions, highlight the inherent difficulty of using a single fixed-parameter potential to capture the composition-driven variation in boron coordination across a wide range of glass chemistries. This limitation is especially pronounced in multi-component systems like borosilicates, where structural adaptation is highly composition-dependent, unlike simpler glasses with a single dominant network former. Despite these limitations, our framework provides the flexibility to navigate such trade-offs effectively. By dynamically adjusting the objective function, we can steer the optimization process toward specific target properties of interest, thereby enabling a tailored balance between competing structural features.

**4.2 Effectiveness of the Proposed Optimization Framework**

Our results highlight the practicality and adaptability of the proposed optimization framework for tuning interatomic potential parameters in complex glass systems. Rather than relying on exhaustive search strategies, the framework leverages a surrogate model with active learning to guide sampling toward the most informative regions of the parameter space. In total, only 400 MD simulations were conducted—230 from the initial dataset and the remaining selected dynamically across just 17 iterations. While we do not benchmark this against a formal baseline, this relatively low simulation count suggests that the framework can reduce computational burden compared to conventional trial-and-error approaches. The framework also supports iterative refinement through dynamic adjustments to the search space, optimization algorithm, and objective function, helping to focus resources on high-potential candidates and improve overall effectiveness.

We obtained two high-quality parameter sets for borosilicate glasses, one from BayOpt and the other from CMA-ES, as summarized in Table 2. Both parameter sets demonstrate strong performance in reproducing density and B⁴ fraction, and both generalized well across a broad range of compositions. Notably, they exhibited different strengths in capturing B⁴ fractions accurately at varying $B_2O_3$ concentrations, reflecting the underlying trade-offs explored in this study. Their agreement with experimental structure factors further supports their physical reliability, as discussed in detail elsewhere[38].

These results demonstrate the effectiveness and flexibility of our composition-aware, multi-objective optimization framework. A key advantage of our approach lies in its direct use of experimental data—rather than simulation-derived targets—which improves physical fidelity and greater robustness to variations in simulation conditions. By jointly optimizing density and B⁴ fraction, the framework balances short- and long-range accuracy, enabling generalization across diverse borosilicate compositions. Compared to brute-force searches, composition-specific parameter fitting, or machine-learned potentials trained on narrow datasets, our method offers a practical compromise between generality, efficiency, and accuracy. Fixed empirical models often miss composition-dependent trends, while composition-specific potentials require complex tuning and lack transferability. Our framework offers a middle ground, providing transferable parameters that generalize well without composition-specific adjustments.

|  | B − O | | | B − B | |
| --- | --- | --- | --- | --- | --- |
|  | $A_{B-O}(eV)$ | $\rho_{B-O}(Å)$ | $C_{B-O}(eV \cdot Å^6)$ | $A_{B-B}(eV)$ | $\rho_{B-B}(Å)$ |
| BayOpt | 4420924.707 | 0.1249337423 | 750.9560382 | 12287.5823 | 0.352685761 |
| CMA-ES | 4607292.953 | 0.1239486039 | 618.6859646 | 11608.31839 | 0.326965741 |

Table 2. Optimized parameter values for BayOpt and CMA-ES Optimization.

Nonetheless, several limitations should be acknowledged. A key challenge lies in the inherent trade-offs observed during optimization—no single parameter set performs equally well across all compositions and structural properties. This is especially true for borosilicate systems, where global density and local $B^4$ coordination are influenced by competing structural factors. While our multi-objective framework explicitly manages these trade-offs, the balance between accuracy and generality remains a central consideration, particularly when extending the method to more chemically or structurally diverse glass systems. In addition, our focus on density and $B^4$ fraction captures only a subset of relevant structural features. Expanding the set of optimization targets to include additional properties—such as angular distributions, coordination environments of other species, or elastic moduli—could improve transferability but would significantly increase the complexity and cost of optimization.

Moreover, although the active learning framework improves data efficiency, it involves components that require manual tuning. Hyperparameters in the surrogate MLP model must often be adjusted as new MD data are added, and the performance of optimization algorithms like BayOpt and CMA-ES is sensitive to initialization and acquisition strategies. To reduce the risk of convergence to local minima, we performed multiple optimization runs per iteration, which improves robustness but adds to computational cost. Future improvements could include automated hyperparameter tuning, adaptive retraining strategies, or uncertainty-aware modeling to further streamline the workflow and improve scalability.

## Conclusions

We have proposed an active-learning-based framework for optimizing classical interatomic potentials for complex glasses directly against experimental data, taking soda-lime borosilicates as an example. By integrating a surrogate ML model with iterative optimization, the method efficiently explored a complex five-dimensional parameter space using only 400 MD simulations across 17 iterations. Two high-quality parameter sets were identified, one from BayOpt and one from CMA-ES, each performing well across a broad composition range.

The framework effectively balances multiple structural objectives, capturing both global (density) and local ($B^4$ fraction) features. It also accommodates composition-dependent challenges, revealing key trade-offs that arise in glass systems with varying network connectivity such as borosilicates. Despite using a fixed-charge potential, the method successfully prioritized specific performance targets through dynamic objective refinement. Validation against experimental X-ray structure factor measurements further confirms the structural fidelity of the optimized potentials. These results demonstrate that composition-aware, experiment-driven optimization offers a practical and generalizable path for developing transferable force fields in complex, multi-component glasses with minimal computational cost. Although demonstrated for glasses, the approach could be broadly used for other materials systems as well to develop interatomic potentials that are aligned to experimental properties.

## CRediT authorship contribution statement

R.C. and K.Y. both led the conceptualization, data generation and analysis, methodology development, validation, software, visualization, and writing of the manuscript. M.M.S. contributed to conceptualization, data generation, methodology, supervision and writing. N. M. A. K contributed to conceptualization, methodology, supervision and writing. J.M. contributed to Methodology and supervision. F.R. contributed to conceptualization, methodology, project administration, supervision and writing. All authors reviewed the manuscript.

## Declaration of competing interest

The authors declare that they have no known competing financial interests or personal relationships that could have appeared to influence the work reported in this paper.

## Acknowledgments

M.M.S. acknowledges support from the European Union (ERC, NewGLASS, 101044664). N. M. A. K acknowledges the support from Alexander von Humboldt foundation, and Google research scholar award. R.C. acknowledges the support from National Science Foundation under Grant No. 1944510.

## Competing interests

The author(s) declare no competing interests.

## Data availability

The example data and code are available for noncommercial use at GitHub: https://github.com/ruoxia-c/Borosilicate-Glass-Potential.

# Supplementary Information

*for*

Optimization of Transferable Interatomic Potentials for Glasses toward Experimental Properties

# S1. Molecular Dynamic Simulation Details

This section provides additional details about the molecular dynamics (MD) simulations. Table S1 lists the partial charges assigned to each element. Table S2 presents the BKS potential parameters used for each atomic pair, excluding the B–O and B–B interactions, which are the targets of our optimization. All listed values are taken from the original Guillot–Sator interatomic potential [1].

**Tabel S1.** Partial charge assigned for each element.

| ELEMENT | PARTIAL CHARGE (E) |
|---|---|
| O | -0.945 |
| SI | 1.89 |
| B | 1.4175 |
| CA | 0.945 |
| NA | 0.4725 |

**Tabel S2.** BKS potential parameters for atomic pairs (excluding B–O and B–B)

| Pair | $A(eV)$ | $\rho(\text{Å})$ | $C(eV \cdot \text{Å}^6)$ |
|---|---|---|---|
| Na-O | 2774279.995 | 0.17 | 0.0 |
| B-Si | 7787.547002 | 0.29 | 0.0 |
| Si-O | 116089.747 | 0.161 | 1067.655354 |
| O-O | 208071.224 | 0.265 | 1962.277725 |
| O-Ca | 3589789.46 | 0.178 | 974.5339388 |
| *-* | 0.0 | 1.0 | 0.0 |

Note. The *-* row defines default parameters applied to all atom pairs not explicitly listed

## S2. Evolution of Parameter Distributions During Optimization

Figures S1 and S2 illustrate how the distribution of the top 10 selected parameter sets evolved across all five parameters over the course of the optimization. Figure S1 presents the changes in parameter distributions from the initial search range to the first expanded range, as discussed in the main text. For clarity, all search ranges are expressed as percentages relative to the original parameter values reported by Wang et al.[2], consistent with the main text. The search range remains fixed during optimization and is only updated manually as part of the strategy. In the Bayesian Optimization iterations, the initial search range was set to [–10%, +10%] for all parameters, except for B–O $\rho$, which was constrained to [–5%, +10%] to maintain stability. However, because many selected parameters sets clustered near the boundaries (particularly for B–O $A$ and B–O $C$), the search range was expanded to [–15%, +15%] for most parameters, while keeping B–O $\rho$ limited to [–5%, +15%] during CMA-ES iterations 1–6.

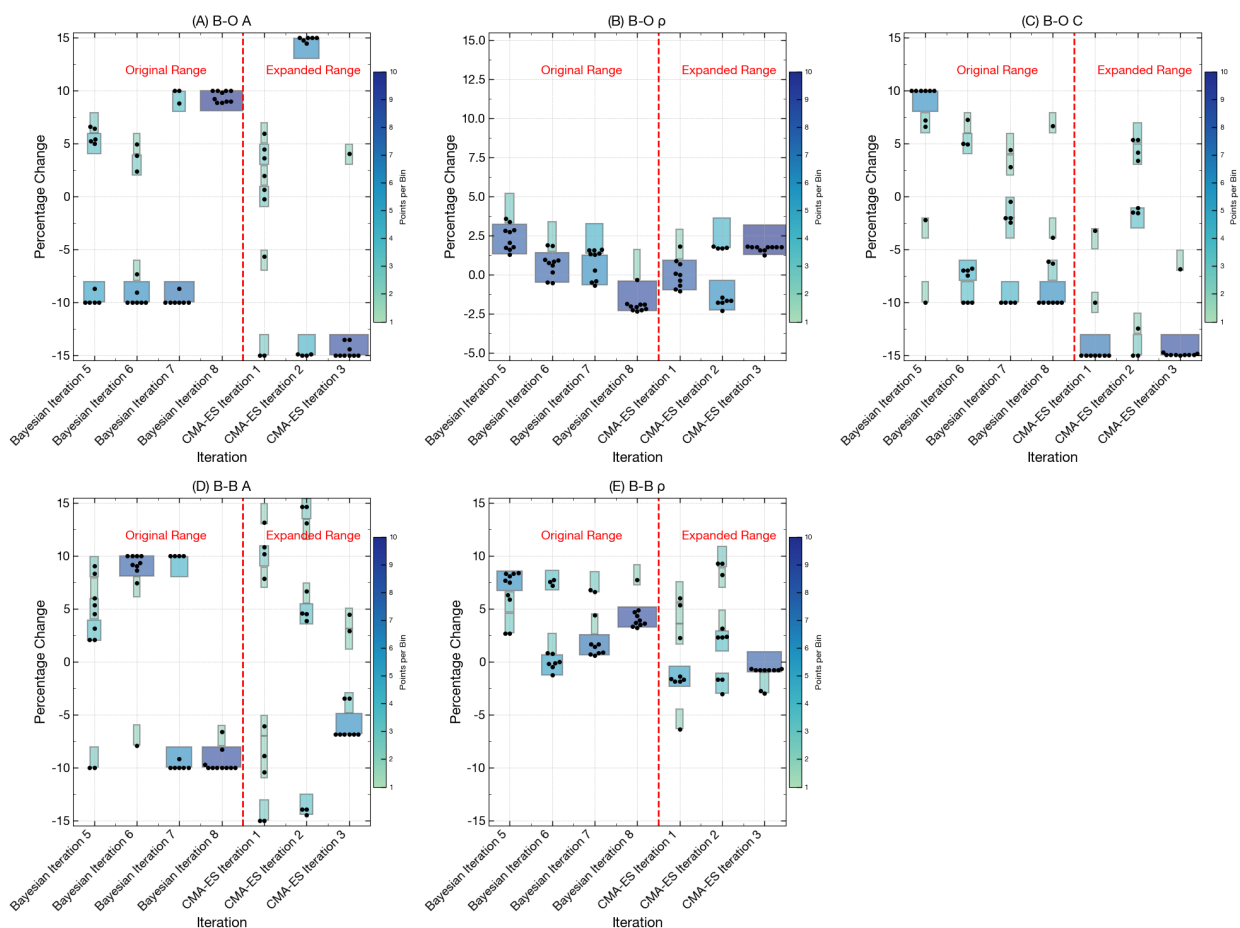

**Figure S1.** Evolution of parameter distributions from initial to first expanded search range during Bayesian Optimization and early CMA-ES iterations.

Figure S2 further shows the parameter distribution trends during the CMA-ES iterations. The left side of the red dashed line corresponds to the first expanded search range, while the right side reflects the second expansion applied in iterations 7 to 9. For these later iterations, the range for B–O *A* was extended further to [–20%, +20%] due to continued boundary clustering. In contrast, B–O *C* values consistently approached only the lower bound, prompting a directional shift to a non-symmetric window of [–25%, +5%] to avoid unnecessary expansion on the upper end. A clear trend is observed, that is, new data points increasingly populate the newly explored parameter regions. Notably, iteration 6 yielded the best-performing parameter set for the CMA-ES optimization, underscoring the efficiency and adaptability of the dynamic optimization strategy employed in this study.

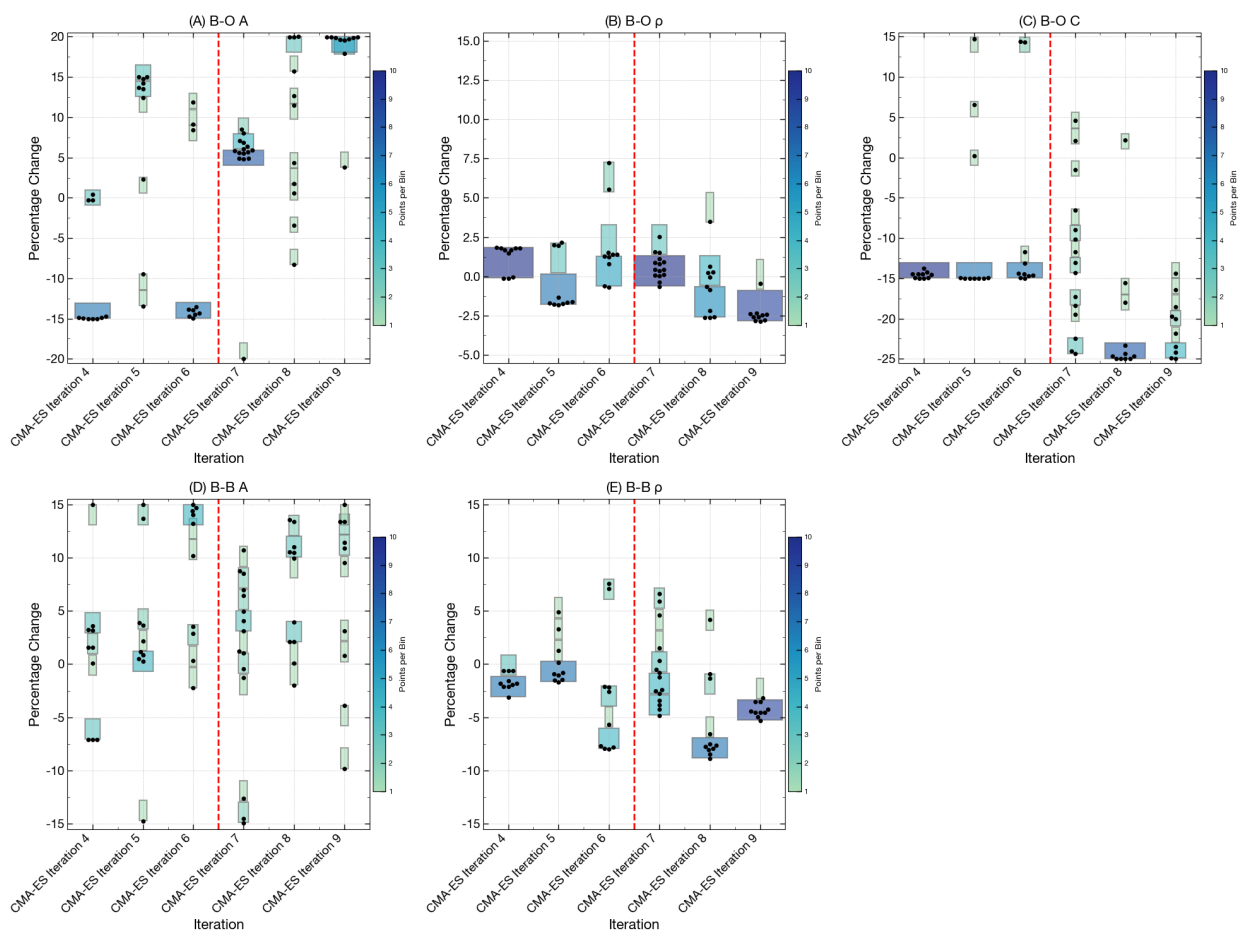

**Figure S2.** Evolution of parameter distributions from first to second expanded search range during CMA-ES iterations 4 to 9.

# S3. Full X-Ray Structure Factor Comparison for Optimized Parameter Sets

To further validate the structural fidelity of the optimized force fields, we present the full set of X-ray structure factor comparisons across all seven borosilicate glass compositions. Figure S3 shows the simulated and experimental structure factors $S(Q)^3$, obtained using the best-performing parameter set from Bayesian Optimization, while Figure S4 presents the corresponding results using the best parameter set from CMA-ES optimization.

Both sets of optimized parameters demonstrate good agreement with experimental measurements, particularly in reproducing the position and intensity of the first sharp diffraction peak, which reflects the medium-range structural ordering of the glass network. While some discrepancies emerge in the high-$Q$ region, especially at higher $B_2O_3$ contents, the overall trends and key features are well captured by both parameterizations. More detailed analysis and discussion of these structural results can be found in ref. [3].

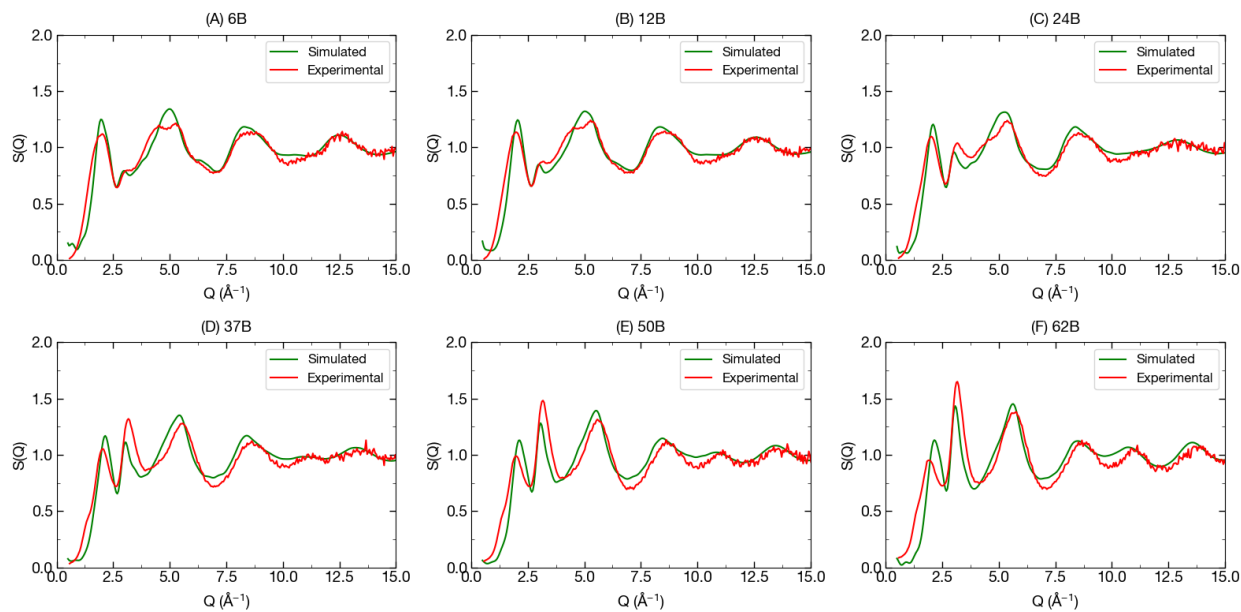

**Figure S3.** Comparison of simulated and experimental X-ray structure factors $S(Q)$ for six borosilicate glass compositions using the best Bayesian Optimization parameter set. Panels (A)–(F) correspond to glasses with increasing $B_2O_3$ content: 6B, 12B, 24B, 37B, 50B, and 62B.

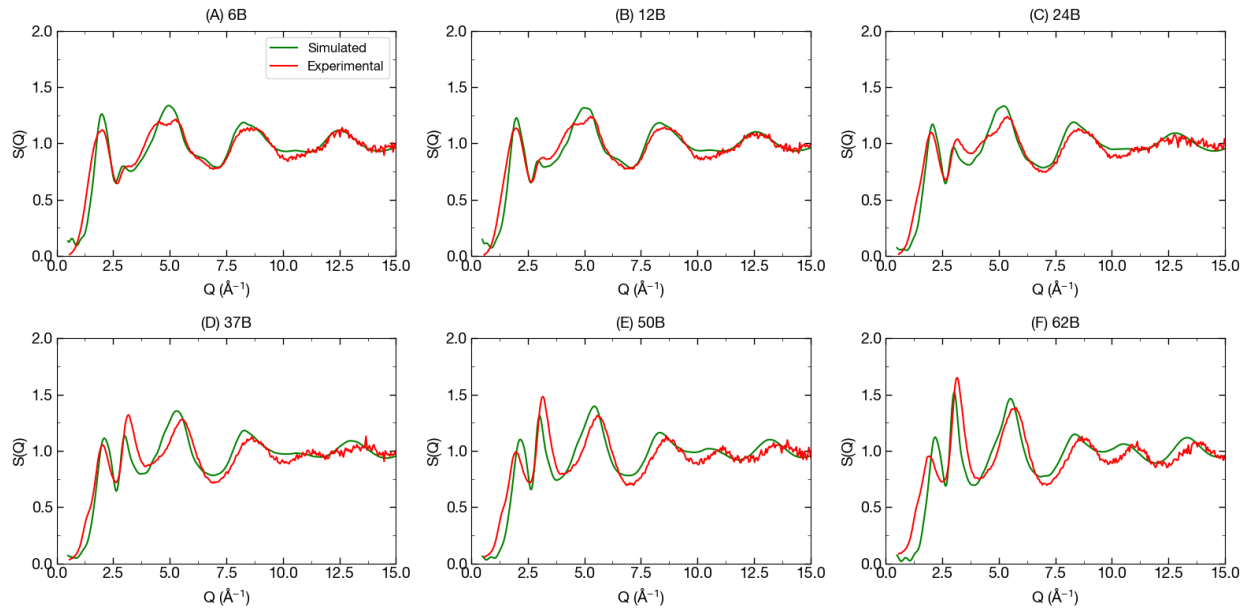

**Figure S4.** Comparison of simulated and experimental X-ray structure factors S(Q) for six borosilicate glass compositions using the best CMA-ES Optimization parameter set. Panels (A)–(F) correspond to glasses with increasing $B_2O_3$ content: 6B, 12B, 24B, 37B, 50B, and 62B.

## S4. Density vs B4 trade-off Discussion

To investigate the trade-off between density and B4 prediction accuracy, we compute the Mean Squared Errors (MSEs) for both properties across all parameters sets tested by MD simulation, including both MD initial dataset and optimized parameter sets. The calculation of MSE follows the same approach described in the main text. Specifically, the normalized density and $B^4$ MSEs are computed according to Equation (4) and Equation (5), respectively.

Figure S5 presents a Pareto front plot that illustrates the trade-off between the two objectives: B4 and density MSE. This type of analysis is commonly used in multi-objective optimization to visualize the balance between competing goals[4]. In the plot, each blue dot represents a tested parameter set, with its corresponding MSE for B4 shown on the x-axis and for density on the y-axis. A solution is considered Pareto optimal, as indicated by the red markers in Figure S5, if no other solution exists that achieves lower error in both objectives at the same time. The collection of Pareto optimal solutions forms a boundary beyond which no further improvement in one objective can be made without increasing the error in the other. To better visualize this trade-off region, the axes are limited to the range [0, 0.05] in both directions.

The plot reveals that all Pareto optimal solutions achieve B4 MSEs close to 0.004, suggesting that B4 prediction is already near its optimal limit under the current modeling framework. Further reducing the B4 error is extremely difficult and often comes at the cost of increased density error. In contrast, the density MSEs among Pareto optimal points vary more widely, which indicates that there is greater potential to reduce density error without significantly affecting B4 performance, up to a certain point. Once density MSE approaches its minimum, further improvement leads to noticeable degradation in B4 accuracy. This pattern highlights the inherent conflict between the two objectives and underscores the value of the Pareto front in guiding the selection of parameter sets based on the desired balance between B4 and density prediction.

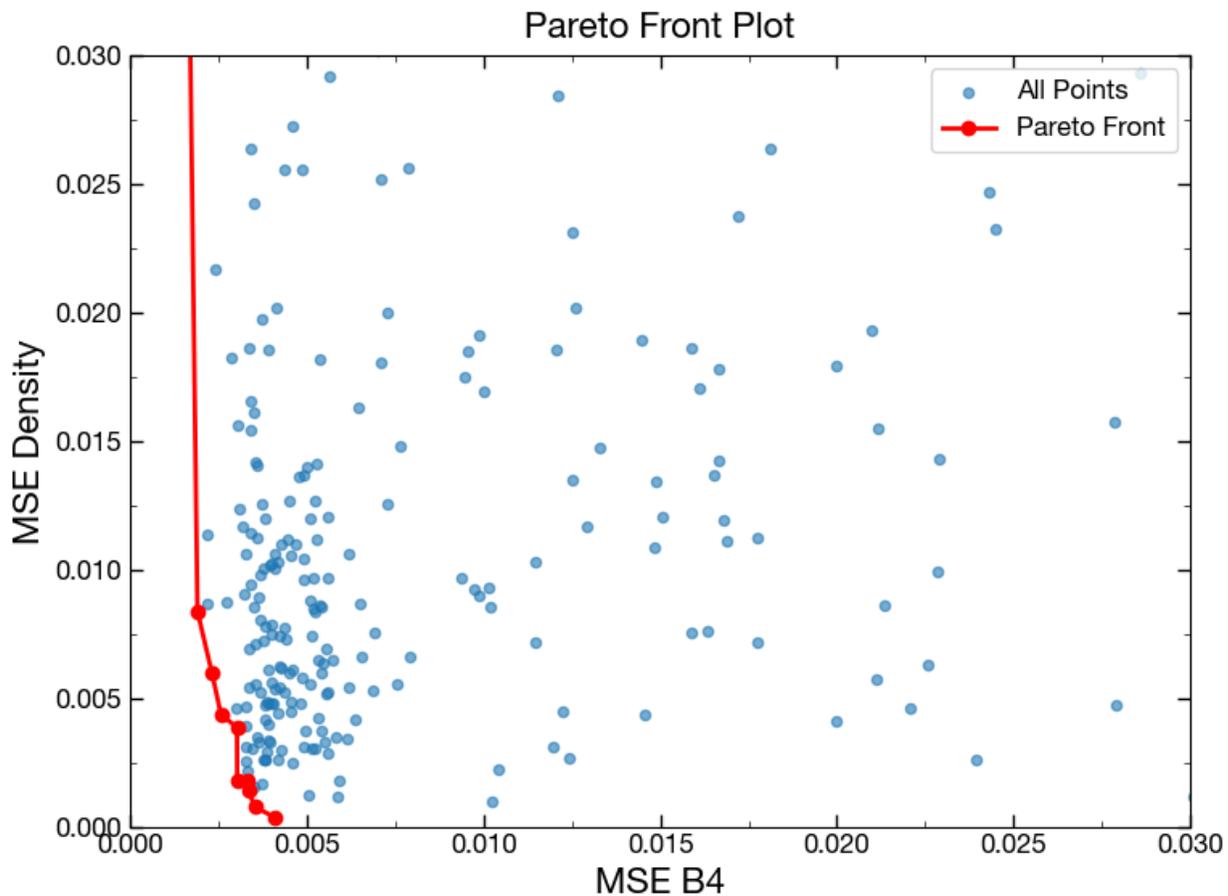

**Figure S5.** Pareto front between B4 and density MSE. Blue dots represent all tested parameter sets. Red markers highlight the Pareto-optimal solutions, where no further improvement can be made in one objective without degrading the other. The plot is zoomed into the [0, 0.05] region on both axes to highlight the most optimal trade-offs.